\documentclass[twocolumn,showpacs,aps,prb,floatfix]{revtex4}
\usepackage{epsfig}

\begin{document}

\title{Electronic structure and magnetic anisotropy of CrO$_2$}
\author{A. Toropova}
\affiliation{Center for Materials Theory, Department of Physics
and Astronomy, Rutgers University, Piscataway, NJ 08854}
\author{G. Kotliar}
\affiliation{Center for Materials Theory, Department of Physics
and Astronomy, Rutgers University, Piscataway, NJ 08854}
\author{S. Y. Savrasov}
\affiliation{Department of Physics, New Jersey Institute of
Technology, Newark, NJ 07102}
\author{V. S. Oudovenko}
\affiliation{Bogoliubov Laboratory for Theoretical Physics, Joint
Institute for Nuclear Research, 141980 Dubna, Russia}
\affiliation{Center for Materials Theory, Department of Physics
and Astronomy, Rutgers
University, Piscataway, NJ 08854}
\date{\today}

\begin{abstract}
The problem of importance of strong correlations for the
electronic structure, transport and magnetic properties of
half--metallic ferromagnetic CrO$_{2}$ is addressed by performing
density functional electronic structure calculations in the local
spin density approximation (LSDA) as well as using the LSDA+U
method. It is shown that the corresponding low--temperature
experimental data are best fitted without accounting for the
Hubbard $U$ corrections. We conclude that the ordered phase of
CrO$_{2}$ is weakly correlated.
\end{abstract}

\maketitle

As a compound with multiple industrial applications and its
unusual half--metallic electronic structure, CrO$_{2}$ has
recently attracted a lot of theoretical~\cite{Sc86, Lewis97,
Korotin98, Mazin99, Laad2001, Craco2003} and
experimental~\cite{Tsu97, Basov99, Sta00, Kun02, Kur03} interest.
The main discussion was centered around the role of strong
correlations for the description of its ferromagnetic phase.
Since Cr in its formal $4+$ valence state has two $3d$ electrons
of $t_{2g}$ symmetry, one would expect manifestation of
correlation effects of the Mott--Hubbard nature. On the other
hand, metallic behavior of spin majority band suggests that
Coulomb interactions of the Hubbard type can be screened
out~\cite{Korotin98}. The comparison with the available
photoemission and optical conductivity data did not make the
situation more clear. One--electron spectra calculated using the
LSDA+U method \cite{basics_LDA+U, anisimov_LDA+U} fit well the
photoemission and inverse photoemission experiments with the
choice of intra--atomic Coulomb and exchange parameters $U=3$~eV
and $J=0.87$~eV~\cite{Korotin98,Tsu97}. This indicates the
importance of strong correlations. Contrary to this result, the
LSDA\ optical conductivity calculations explain experimental data
\cite{Mazin99}, which suggests the regime of weak coupling.

In the present paper we address the issue of controversial role
of strong correlations in ferromagnetic CrO$_{2}$ by presenting
combined studies of its electronic structure, optical
conductivity and magnetic anisotropy\ using the LSDA and LSDA$+$U
schemes. We employ a linear--muffin--tin--orbital (LMTO) method
in its atomic sphere approximation (ASA) \cite{OA75,Sav} for our
electronic structure calculations. The low symmetry of the rutile
structure and small packing factor of the unit cell require an
introduction of additional empty spheres. Their positions are
chosen to be $4c$ and $4g$ in Wyckoff notations. The radii of the
spheres (in atomic units) for Cr and O atoms, as well as of the
empty spheres are chosen to be 1.975, 1.615, 1.378 and 1.434,
correspondingly. The basis set adopted in the calculations is
Cr(4s, 4p, 3d) and O(2s, 2p).

In rutile structure Cr atoms are surrounded by distorted oxygen
octahedra. The positions of the octahedra lead to a new natural
basis for Cr orbitals. In this basis the cubic component of the
octahedral crystal field splits the fivefold degenerate $3d$
orbital into higher energy doubly degenerate $e_{g}$ level and
lower energy triple degenerate $t_{2g}$ level. Distortions of
oxygen octahedra further split the $t_{2g}$ states into lower
energy $ t_{2g}^{\parallel }$ orbital ($xy$ character)  and higher
energy twofold degenerate $ t_{2g}^{\perp }$ orbitals ($yz+zx$ and
$yz-zx$ characters)~\cite{Korotin98}.

\begin{figure}[h]
\includegraphics[width=0.4\textwidth]{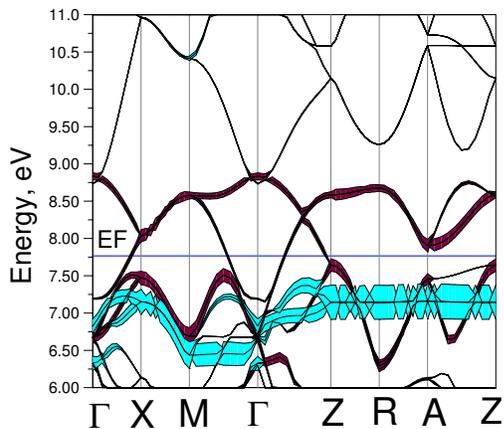}\\
\caption{LSDA band structure of CrO$_{2}$ \ for spin majority
carriers. Dark and light shaded areas show the specific weight of
$t_{2g}^{\perp }$ and $ t_{2g}^{\parallel }$ orbitals
respectively in the particular band.}
\label{fig:CrO2_Majority_band}
\end{figure}

\begin{figure}[h]
\includegraphics[width=0.4\textwidth]{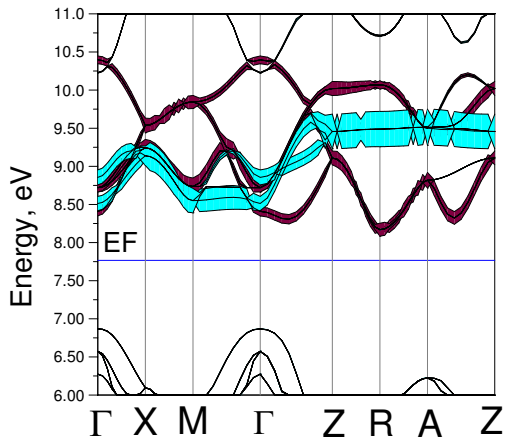}\\
\caption{LSDA band structure of CrO$_{2}$ \ for spin minority
carriers. Dark and light shaded areas show the specific weight of
$t_{2g}^{\perp }$ and $t_{2g}^{\parallel }$ orbitals respectively
in the particular band.} \label{fig:CrO2_Minority_band}
\end{figure}

The results of the LSDA band structure calculation in the
vicinity of the Fermi energy are shown in
Figs.~\ref{fig:CrO2_Majority_band} and
\ref{fig:CrO2_Minority_band}. The Fermi level crosses the spin
majority $t_{2g}$ manifold. The rest of the Cr $3d$--states is
formed from four $e_{g}$--bands and three $t_{2g}$ spin minority
bands which are located above the Fermi level. In both spin
channels $e_{g}$ and $t_{2g}$ bands are well separated for all
momenta except for the $\Gamma $-point. The whole $3d$--complex
is  strongly hybridized with oxygen.
In Fig.~\ref{fig:CrO2_Minority_band} one can see that in the spin
minority channel there is gap of approximately $1.3$~eV between
the oxygen $2p$--band and the chromium $d$--band. This gap leads
to $100\%$ spin polarization at $ E_{F}$ and assures the magnetic
moment to be precisely equal to $4$ $\mu _{B}$ per unit cell. The
$t_{2g}$ bands that cross the Fermi level in the spin majority
channel mainly consist of the $t_{2g}^{\perp }$ orbitals (see
Fig.~ \ref{fig:CrO2_Majority_band}). Almost non--dispersive
narrow band below $ E_{F}$ (shown as lightly shaded) is formed by
the $t_{2g}^{\parallel }$ orbital. This localized state undergoes
large exchange splitting $\Delta ^{ex}$ making spin minority
$t_{2g}^{\parallel }$ orbitals unoccupied (see
Fig.~\ref{fig:CrO2_Minority_band}).

\begin{figure}[h]
\includegraphics[width=0.4\textwidth]{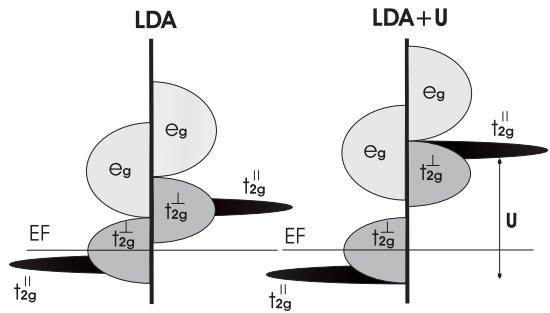}
\caption{Schematic density of states (DOS) of CrO$_{2}$ \ deduced
from the LSDA and LSDA$+$U calculations. Shaded semicircles from
right and left represent the bands for spin majority and spin
minority carriers.} \label{fig:levels}
\end{figure}

The main changes which occur in the band structure for non--zero
values of $ U $ and $J$ using the LSDA$+$U method are
schematically shown in Fig.~\ref{fig:levels}. These calculations
were performed with $U=3$~eV and $ J=0.87 $~eV. The center of
gravity of occupied $t_{2g}^{\parallel }$ band is pushed down by
$0.6$~eV. The spin minority unoccupied $e_{g}$ bands are pushed
up by $0.6$~eV, which opens  $0.4$~eV gap between $t_{2g}^{\perp
}$ and $e_{g}$ bands above the Fermi level. In the spin minority
channel the occupied oxygen bands are shifted up by $0.3$~eV. The
upper unoccupied $t_{2g} $ and $e_{g}$ bands are shifted up by
$1.1$~eV. As a result, the insulating gap is increased and
reaches the value of $2.1$~eV.

\begin{figure}[h]
\includegraphics[width=0.3\textwidth,angle=-90]{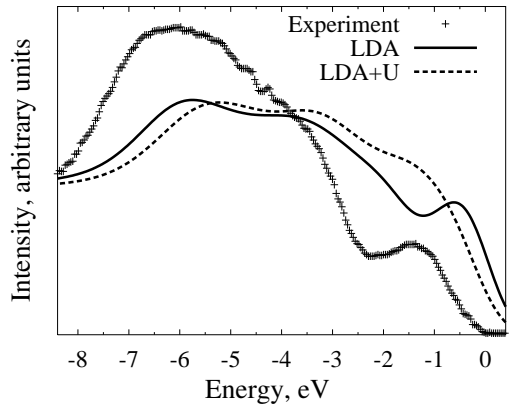}%
\vspace{0.5cm} \caption{Comparison between theoretical densities
of states and experimental \protect\cite{Tsu97} UPS spectra for
CrO$_{2}$. The theoretical DOS were smeared out by Gaussian and
Lorentzian broadening functions to account for experimental
resolution and lifetime effects. The secondary electron
background has been taken into account. } \label{fig:UPS}
\end{figure}

Now we compare our calculated electronic structure using the LSDA
and the LSDA+U method with the available experimental data.
Fig.~\ref{fig:UPS} shows comparison of ultraviolet photoemission
spectroscopy (UPS) experiments~\cite{Tsu97} (photon energy $h\nu
=40.8$~eV) with the theoretical spectra which are calculated
densities of states smeared by both Gaussian and Lorentzian
broadening functions. The Gaussian broadening takes into account
experimental resolution while Lorentzian takes into account
finite lifetime effects. The Gaussian broadening parameter is
taken to be $0.4$~eV. The full width at half maximum (FWHM) of
the Lorentzian was taken to be energy dependent and equal to
$0.2|E-E_{F}|$~eV. We can distinguish two main features in the
UPS spectra: (i) a small hump in around $-1.5$~eV which arises
from the $t_{2g}$ band of Cr, and (ii) a big hump around
$-6.0$~eV which comes from the broad $2p$ oxygen band. Both
features are fairly well described by both the LSDA and the
LSDA$+$U calculation. The small discrepancy between the LSDA
calculation and experiment could be referred to the fact that at
small photon energies photoemission is a more surface sensitive
technique. Indeed, recent PES studies of Vanadium
oxides~\cite{Mo03} have been found to yield spectra not
characteristic of the bulk, but rather of surface atoms whose
lower coordination number can render more strongly correlated
surface layer.

\begin{figure}[tbh]
\includegraphics[width=0.3\textwidth,angle=-90]{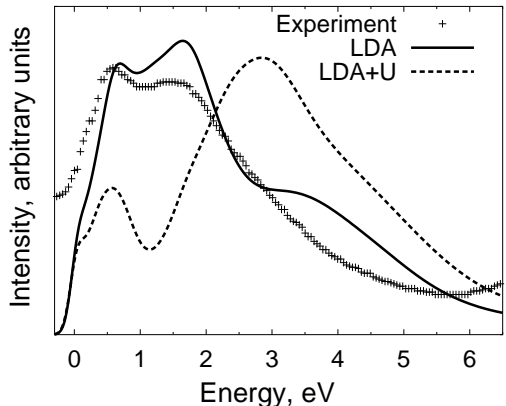}
\caption{Comparison between theory and
experiment\protect\cite{Kur03} for Cr 2p x-ray absorption (XAS)
spectrum. To deduce theoretical curve from the partial Cr 3d DOS
we used $0.1$~eV for Gaussian FWHM. The Lorentzian FWHM was taken
to be energy dependent and equal to $0.2|E-E_{F}|$. The binding
energy of core $2p_{3/2}$ Cr state 577~eV has been subtracted
from the experimental spectrum.} \label{fig:XAS}
\end{figure}

For the unoccupied states we have chosen to compare our results
with the available x--ray absorption spectra (XAS) \cite{Kur03}
rather than with the inverse photoemission as it had been done
before~\cite{Tsu97}. The main reason for this is that XAS is a
bulk (not surface) sensitive method. The $2p $ Cr XAS
spectrum~\cite{Kur03} is compared to our theoretical calculations
in Fig.~\ref{fig:XAS}. To deduce theoretical spectra we performed
both Gaussian and Lorentzian broadening of $3d$ and $4s$ partial
DOSes. Two first peaks around $0.5$~eV and around $1.5$~eV come
from the unoccupied $3d$ orbitals of chromium. The main
contribution to the second peak comes from the $t_{2g}$ orbitals
in the spin minority channel. Thus, the LSDA+U overestimates the
spin minority gap twice as much.

\begin{figure}[h]
\includegraphics[width=0.3\textwidth,angle=-90]{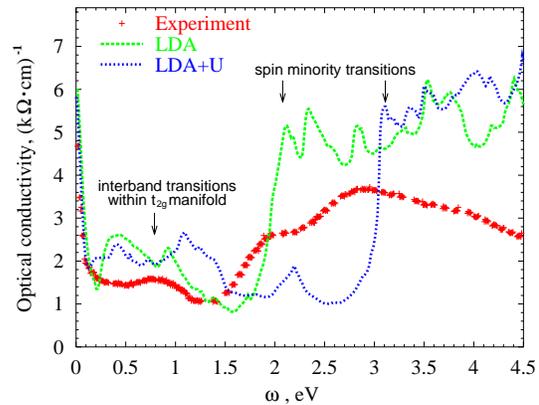}
\caption{Comparison of the optical conductivity of CrO$_{2}$
obtained using the LSDA and LSDA+U methods against the
experimental data \protect\cite{Basov99}.} \label{fig:OPT}
\end{figure}

Below we discuss the optical conductivity of CrO$_{2}$. In
Fig.~\ref{fig:OPT} diagonal $x$-components of the optical
conductivity calculated using the LSDA and LSDA+U methods are
compared with the experimental results reported by Basov and
coworkers ~\cite{Basov99} ($x$ coordinate refers to the basis of
unit cell). The main two features of the calculated optical
conductivity are a shoulder around $2-3$~eV and a broad hump
located at energies $0.2-1.5$~eV. In both LSDA and LSDA+U schemes
the shoulder can be identified with two types of transitions.
First contribution arises from the minority spin gap transitions
and the second one comes from transitions between the occupied
$t_{2g}^{\parallel }$ and unoccupied ${e_{g}}$ bands. The hump is
formed by interband transitions within the $t_{2g}$-manifold and
the oxygen $2p$ bands near the Fermi level in the spin majority
channel. Apparently, the LSDA prediction is much closer to the
experimental curve than the LSDA+U one. The LSDA+U calculations
overestimate the minority gap, and due to that, the spin minority
transitions occur at higher energies.

Results of the calculated magnetic anisotropy of CrO$_{2}$ are
presented below. Magnetic anisotropy is the dependence of internal
energy on the direction of spontaneous magnetization. The magnetic
anisotropy is a relativistic phenomenon arising due to
spin--orbit coupling, where the spin degrees of freedom interact
with the spatial anisotropy through the coupling to the orbital
degrees of freedom. This induces a preferred direction of spins.
Because CrO$_{2}$ is a metastable compound, which irreversibly
decomposes at about 200~$^{\circ }$C, all measurements for this
material are performed on polycrystals, microgranulus or thin
films. The first reliable result on magnetocrystalline anisotropy
measurements was reported in Ref.~\onlinecite{Clo62}. The
discovery of the atmospheric pressure chemical vapor
decomposition (CVD) technique has allowed to grow high--quality
films of CrO$_{2}$. As a result, in recent years a lot of studies
of magnetic properties were performed on epitaxial CrO$_{2}$
layers deposited on single crystal (100) TiO$_{2}$ substrates
\cite{Sp00,Ya00,Li99}. For thicker films (700~\AA -1.2$\mu m$)
the in--plane magnetic anisotropy was observed with [001] and
[010] easy and hard axis directions respectively. The
magnetocrystalline anisotropy constant $K_{1}$ has been reported
by different groups to be $ 4.4\times 10^{5}erg/cm^{3}$
~~\cite{Li99}, $2.7\times 10^{5}erg/cm^{3}$ ~~\cite{Ya00} and
$1.9\times 10^{5}erg/cm^{3}$ ~~\cite{Sp00}. However, these values
can significantly differ from the bulk quantities because of a
large lattice mismatch between CrO$_{2}$ films and TiO$_{2}$
substrates (till $4\%$ ). The relaxation as a function of
thickness is very gradual and even for $ 1.2\mu m$ films magnetic
anisotropy shows significant deviation from the bulk value.

We calculate the magnetic anisotropy energy (MAE) by taking the
difference of the two total energies with different directions of
magnetization ([001], [010], [111] and [102]). For the momentum
space integration, we follow the analysis given by Trygg and
co--workers~\cite{tjew} and use special point
method~\cite{froyen} with a Gaussian broadening~\cite{mp} of $15\
mRy$. The validity and convergence of this procedure has been
tested in their work ~\cite{tjew}. We used about $1000\
\mathbf{k}$--points in the irreducible Brillouin zone, while the
convergence of MAE is tested up to $8000$ $\mathbf{k}$--points.

The direction [001] was found to be easy magnetization axis
within our LSDA calculation which is consistent with latest thin
film experiments \cite{Sp00,Ya00,Li99}. Numerical values of MAE
in this case exceed the maximum experimental value by
approximately two times~\cite{Li99}.

\begin{figure}[h]
\includegraphics[width=150pt,angle=-90]{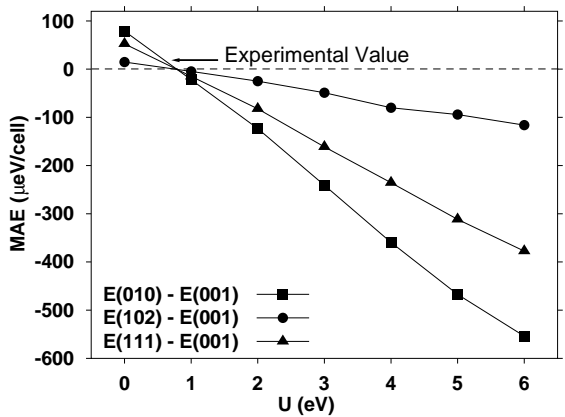}
\caption{The magneto--crystalline anisotropy energies for CrO$ _2
$ \ as functions of $U$. The experimental value of MAE $
E[010]-E[001]=15.6 \protect\mu $~eV per cell is shown by arrow.}
\label{fig:MAE_vs_U}
\end{figure}

To figure out the influence of intra--atomic repulsion $U$ on the
magnetic anisotropy, we have performed LSDA$+$U calculations for
different values of $U$ increasing it from 0 to $6$~eV
($J=0.87$~eV has been kept constant except for the LSDA $U=0$
case). The results of these calculations are presented in
Fig.~\ref{fig:MAE_vs_U}. MAE decreases rapidly starting from the
LSDA value (which is approximately equal to $68$ $\mu$~eV per
cell) and changes its sign around $U\approx 0.9$~eV. This leads to
switching correct easy magnetization axis [001] to the wrong one,
namely [102]. The biggest experimental value of the MAE reported
in the literature is $15.6$ $ \mu $~eV per cell~\cite{Li99}. The
calculated MAE approaches this value around $U=0.6$~eV. This
signals that correlation effects in the $d$--shell may be
important for this compound although they are strongly screened
out.

To conclude, we have reported the LSDA and LSDA$+$U calculations
of electronic structure, optical conductivity and magnetic
anisotropy of CrO$_{2}$. Our comparisons with the experimental
data revealed that the best match is provided by the local spin
density approximation. We explained the discrepancy between the
LSDA and photoemission studies, discussed earlier by other
authors~\cite{Korotin98,Tsu97}, by the fact that due to small
photon energies used in PES, it is more surface rather than bulk
sensitive technique. We resolved this problem by showing that XAS
spectrum is unambiguously described by the LSDA calculation. Is
has been also shown that even intermediate  values of $U$(~of the
order of 1-2~eV) lead to the failure of the LSDA+U method to
describe the magnetic anisotropy and the optical conductivity of
CrO$_{2}$. Since the LSDA$+$U is not adequate for the description
of electronic structure of CrO$_{2}$ \ as well as of its optical
and magnetic properties, we conclude that the ordered phase of
CrO$_{2}$ could be described as weakly correlated material with
small values of on-site Coulomb repulsion. It is important to
notice that,  while we have found that the a simple one-electron
picture describes well the ferromagnetic phase of this material,
there is a narrow band formed by the non-dispersive $t_{2g}^{||}$
orbitals ($xy$ character) which in the paramagnetic phase will be
single occupied, due to the on-site Coulomb interactions, an
effect which cannot be described in LDA and will require a
Dynamical Mean-Field treatment for this materials as done in
Ref.~\onlinecite{Laad2001}. The physical basis for the
applicability of static mean--field picture in the {\it
ferromagnetic phase} of this material, is due to the large
exchange splitting which is able to effectively enforce the
single occupancy of the $t_{2g}^{||}$ orbitals.

This research was supported by the ONR grant No.~4-2650. The
authors would like to thank I. I. Mazin for helpful and
enlightening discussions.

\end{document}